\documentclass[12pt,journal,onecolumn]{IEEEtran}
\usepackage{cite}

\ifCLASSINFOpdf
  \usepackage[pdftex]{graphicx}
  \DeclareGraphicsExtensions{.pdf,.jpeg,.png}
\else
  \usepackage[dvips]{graphicx}
  \DeclareGraphicsExtensions{.eps}
\fi
\usepackage[cmex10]{amsmath}
\usepackage{amssymb}
\usepackage[tight,footnotesize]{subfigure}

\usepackage[caption=false,font=footnotesize]{subfig}
\usepackage{fixltx2e}

\usepackage{stfloats}

\ifCLASSOPTIONcaptionsoff
  \usepackage[nomarkers]{endfloat}
 \let\MYoriglatexcaption\caption
 \renewcommand{\caption}[2][\relax]{\MYoriglatexcaption[#2]{#2}}
\fi

\begin{document}
%
\title{Channel Sounding Waveforms Design for Asynchronous Multiuser MIMO Systems}
\author{Zhenhua Yu$^{\star}$, Robert J. Baxley$^{\dagger}$, Brett T. Walkenhorst$^{\dagger}$, and G. Tong Zhou$^{\star}$\\
$^{\star}$ School of Electrical and Computer Engineering, Georgia Tech, Atlanta, GA 30332-0250, USA\\
$^{\dagger}$ Georgia Tech Research Institute, Atlanta, GA 30332-0821, USA}

\maketitle
\begin{abstract}
In this paper we provide three contributions to the field of channel sounding waveform design in asynchronous Multi-user (MU) MIMO systems. The first contribution is a derivation of the asynchronous MU-MIMO model and the conditions that the sounding waveform must meet to independently resolve all of the spatial channel responses. Next we propose a chirp waveform that meets the constraints and we show that the MSE of our system meets the Cramer-Rao Bound (CRB) when the time offset is an integer multiple of the sampling interval. Finally we demonstrate that the channel capacity region of the asynchronous system and synchronous system is equivalent under certain conditions. Simulation results are provided to illustrate the findings.
\end{abstract}
%
\begin{IEEEkeywords}
Multi-user, MIMO, Asynchronous, Channel sounding, chirp waveform
\end{IEEEkeywords}

%
\IEEEpeerreviewmaketitle

\section{Introduction}
%
%
%
%

\IEEEPARstart{M}{ultiple} input multiple output (MIMO) communication is a powerful way to increase the channel capacity and combat the effects of fading in wireless communication systems\cite{Foschini1998,Telatar1999}. The channel state information (CSI) is very crucial for data detection and channel equalization in MIMO systems, as well as for determining the capacity of a given MIMO deployment configuration. Multi-user MIMO (MU-MIMO) is being deployed in both the LTE advanced and WiMax standards \cite{Li2010, MAN} as a method for spatially multiplexing multiple user data streams in the same time/frequency slot in order to maximize channel utilization.

In order to ascertain the capacity and hence the effectiveness of MU-MIMO, it is crucial to perform channel sounding tests to determine the spatial channel response across all of the MIMO channels. Channel sounding is much more complicated in MU-MIMO systems when trying to sound the channel from multiple sources that have significant distance separation. Furthermore, one or all of the users may be moving relative to each other.

Most of the MIMO channel sounders \cite{Lehne1998,Steinbauer2000,Beach2000,Wirnitzer2001} simply extend the single input single output (SISO) channel sounders, which use a multiplexing switch to scan the different antennas sequentially. The sequential techniques avoid the cost and complexity of the multiple receiver architecture. The works in \cite{Salous2002,Gokalp2002,Salous2005,Salous2010} proposed a semi-sequential architecture by using parallel receivers; however, the switching is still expected at the transmitter side. Any system that uses switching will be unable to capture true cross channel statistics estimates, which are required in order to estimate channel capacity and other relevant channel metrics.  This shortcoming is usually ignored, or explained away by stating that the channel coherence time is longer than the time it takes to cycle through all channels, but as the size of MIMO arrays get larger, such coherence-time constraints become increasingly unrealistic.  Another drawback to systems that have to cycle through antennas is that they are limited in the Doppler resolution that can be resolved.  Doppler resolution is inversely proportional to the channel sounding waveform repetition rate.  In switching systems, this repetition rate increases with the number of antennas inherently limiting the Doppler bandwidth that can be resolved.  In this paper, we propose a code-base sounding waveform that enables both increased Doppler resolving power as well as the ability to estimate channel cross correlation statistics.

Ideally, each of the transmitters and receivers would be time-synchronized to the same atomic clock timer.  Under certain test scenarios, this may be feasible and even straightforward to accomplish.  However, given that channel sounding is usually performed with signal generators at the transmitter and signal analyzers at the receiver, there is no easy duplex channel to facilitate clock synchronization. GPS synchronization may be used, but is not available for indoor channel sounding.  Of course, even when it is possible to perform synchronization across all nodes, this will come at the cost of increased expense and complexity.

This paper addresses waveform design for channel sounding in such an asynchronous MU-MIMO system.  We provide the following contributions: i) We derive the asynchronous MU-MIMO model and we show the conditions that the sounding waveform must meet in order to independently resolve all of the spatial channel responses. ii) We propose a chirp waveform that meets the constraints. We find the Cramer-Rao Bound (CRB) for any channel sounding system and then show that the MSE of our system meets the CRB when the time offset is an integer multiple of the sampling interval. For fractional time offset, the performance is degraded by an unknown pulse shaping matrix; iii) We demonstrate that the capacity is invariant to the channel collected via asynchronous collection versus perfectly synchronous collection as long as either the transmitter side or the receiver side is synchronized.

\section{System Model}

\begin{figure}[!t]
\centering
\includegraphics[width=3.6in]{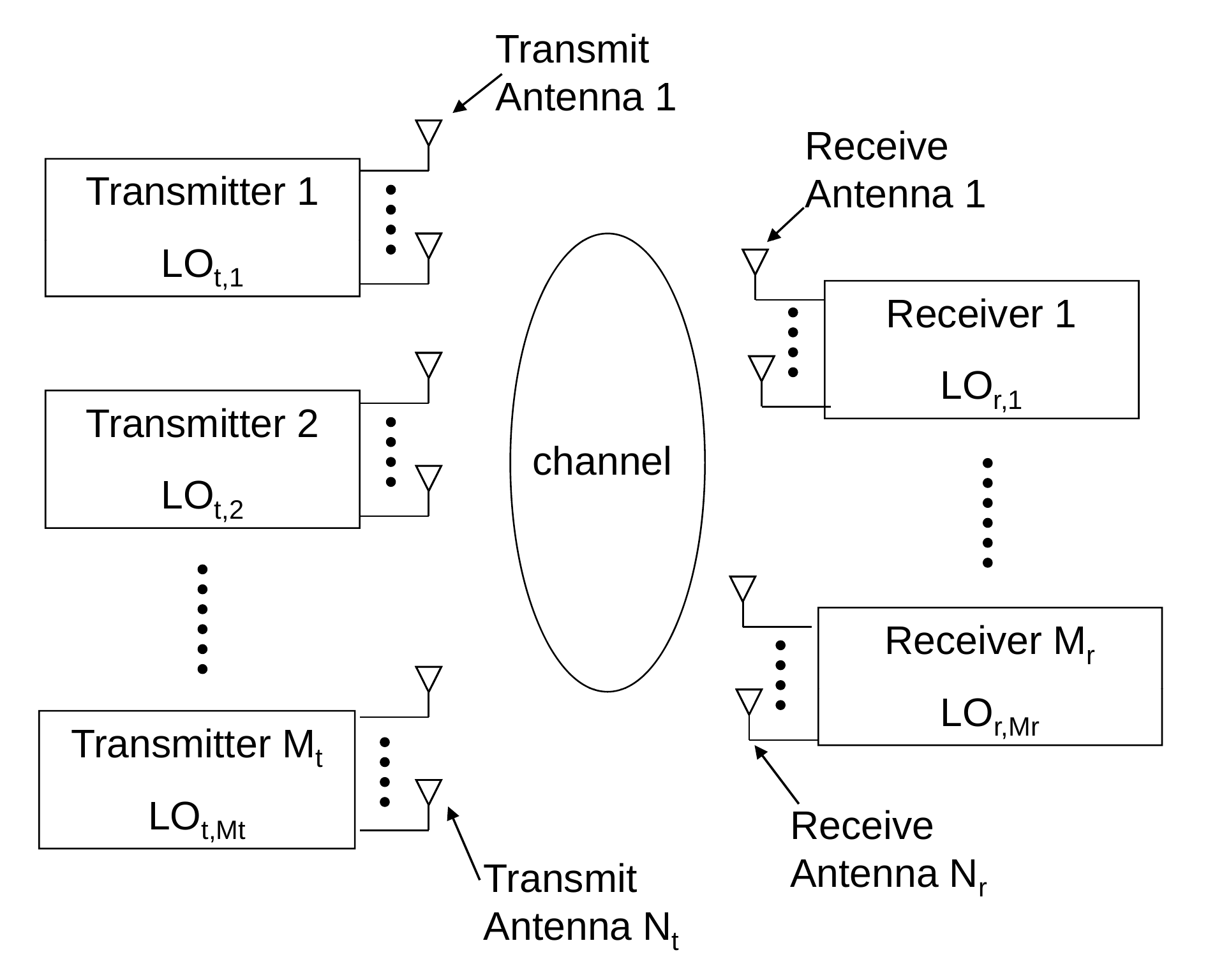}
\caption{Multi-user MIMO systems}
\label{fig_MUMIMO}
\end{figure}

We consider an MU-MIMO system with $M_t$ transmit nodes and $M_r$ receive nodes. Assume $N_t$ and $N_r$ to be the total number of antennas for all the transmit nodes and all the receive nodes respectively, then the system has $N_t\times N_r$ channels to be sounded, as illustrated in Fig. \ref{fig_MUMIMO}. Each channel is assumed to be frequency-selective and block-wise time invariant over the duration of the sounding waveform. We denote the channel response by $h_{i,m}[l]$ for the multipath channel response between the $i$th transmit antenna and the $m$th receive antenna at lag $l$, where $i=1,2,\dots,N_t,\,m=1,2,\dots,N_r$, and $l=0,1,\dots,L-1$. The continuous clock mismatch delay of the link between the $i^{th}$ transmit antenna and the $m^{th}$ receive antenna is denoted by $\zeta_{i,m}$. In the asynchronous multi-user system, each transmitter and receiver has one local oscillator (LO) which is not synchronized between nodes; thus the clock mismatch delays will be equivalent if transmitter and receiver antennas belong to the same transmitter-receiver pair but are different otherwise. Let $T$ denote the sampling interval. By dividing the clock mismatch delay $\zeta_{i,m}$ by $T$, we can obtain
\begin{equation}
\frac{\zeta_{i,m}}{T}= d_{i,m} + \mu_{i,m},
\end{equation}
where $d_{i,m}$ denotes the integer part and $\mu_{i,m} \in [-\frac{1}{2}, \frac{1}{2}]$ denotes the fractional part. Without loss of generality, we let $\mu_{i,m}$ be positive in this paper; i.e., $\mu_{i,m} \in (0, \frac{1}{2}]$. In the following, we will use parentheses $(\cdot)$ to represent the continuous-time function and use square brackets $[\cdot]$ to represent discrete-time function.

\subsection{With only integer offset}
\begin{figure}[!t]
\begin{minipage}[b]{1.0\linewidth}
  \centering
  \centerline{\includegraphics[width=12cm]{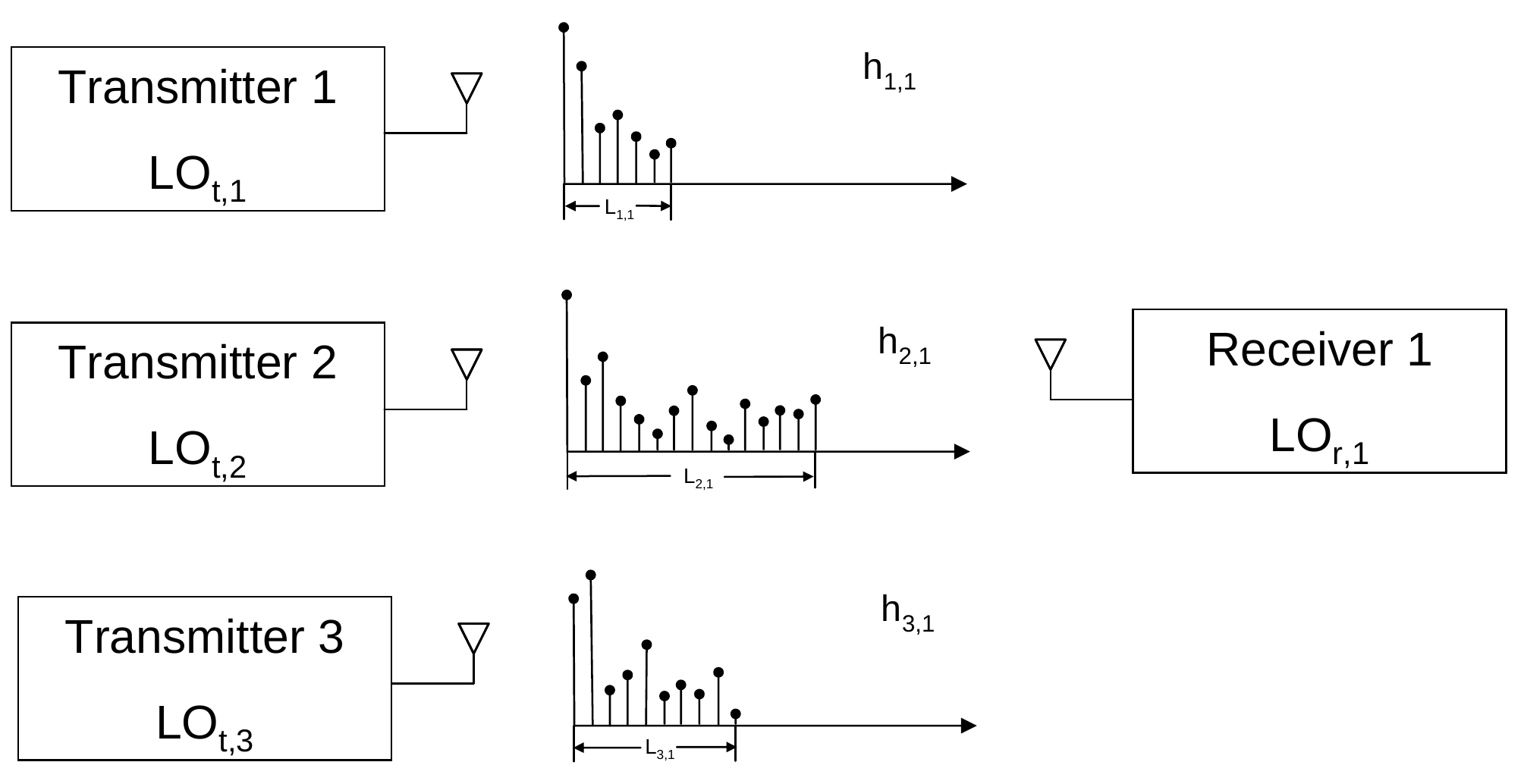}}
\centerline{(a) Synchronous transmissions}
\end{minipage}
\hfil

\begin{minipage}[b]{1.0\linewidth}
  \centering
  \centerline{\includegraphics[width=12cm]{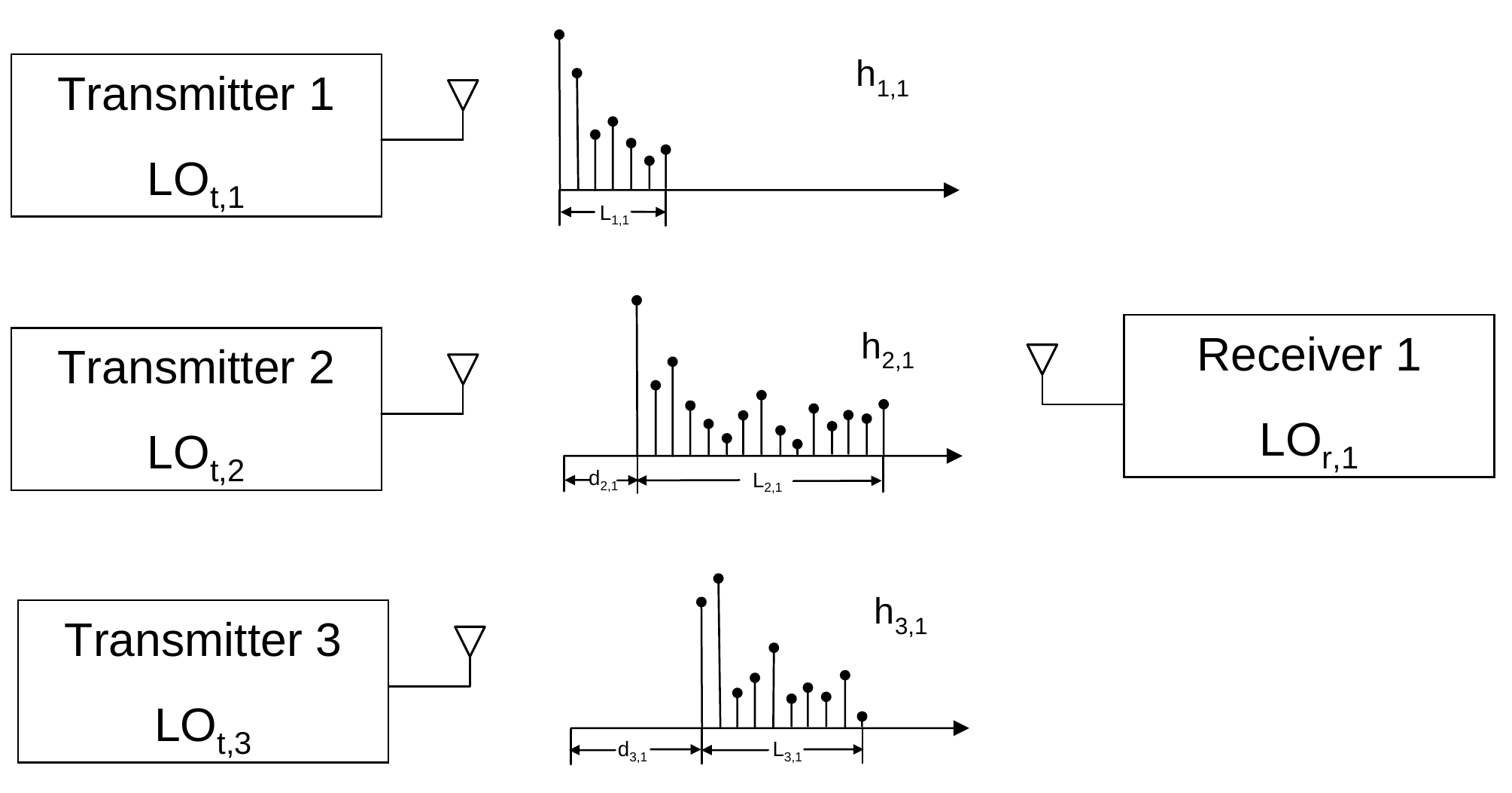}}
\centerline{(b) Asynchronous transmissions}
\end{minipage}
\caption{Illustrations of Multi-user MIMO channels with synchronous and asynchronous transmissions}
\label{fig:channel}
\end{figure}

We first consider the scenario with only integer offset. We model the effects of integer offset as part of the channel response. The first $d_{i,m}$ taps of the channel response are considered to be zeros. Without loss of generality, we assume that $d_{1,m}=0, m=1,\dots,N_r$. Assume that the number of nonzero taps of the channel response between the $i^{th}$ transmit antenna and the $m^{th}$ receive antenna is $L_{i,m}$ and $L_{i,m}+d_{i,m}\leq L$. Two examples of MU-MIMO channels with synchronous and asynchronous transmissions, respectively, are illustrated in Fig. \ref{fig:channel}, where $M_t = 3, N_t = 3, M_r =1, N_r = 1$.


\begin{figure}[!t]
\centering
\includegraphics[width=6.7in]{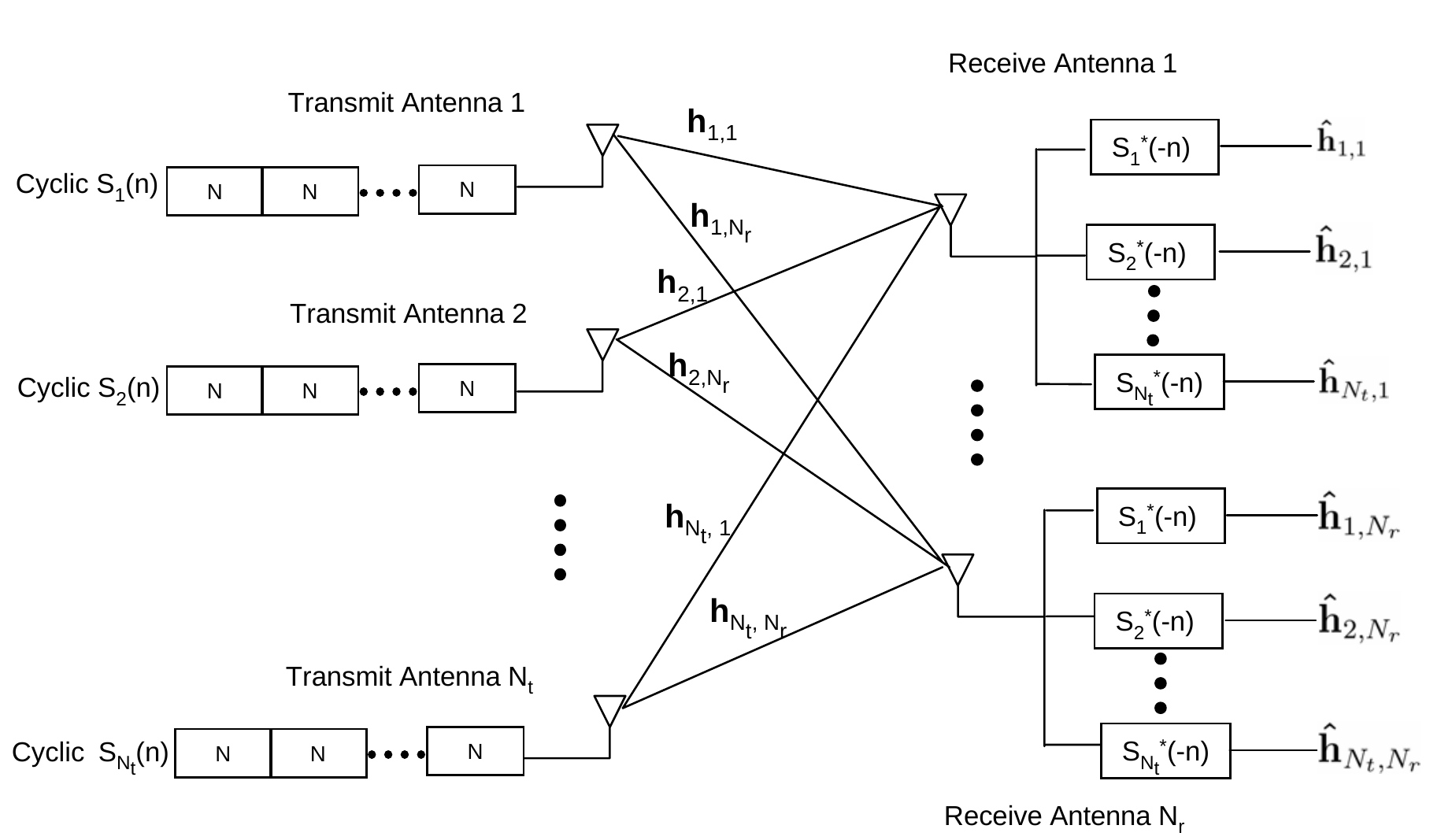}
\caption{Illustration of channel sounding in Multi-user MIMO systems}
\label{fig_egsounding}
\end{figure}

Each transmit antenna of the channel sounder sends out a signal that consists of periodically repeated sequences \cite{Molisch2011}. Let $\mathbf{s_i}$ be the cyclic sounding waveform with period N transmitted by the $i^{th}$ antenna. The signal at the receive antenna $m$ at time $n$ can be expressed as
\begin{equation}
\label{model_1}
r_m[n]= \sum_{i=1}^{N_t}\sum_{k=0}^{L-1}h_{i,m}[n]s_i[n-k]+z_m[n],\,\,n=0,\dots,N-1,
\end{equation}
where $z_m[n]$ is additive noise at the receive antenna $m$ at time $n$ assumed to be white Gaussian (AWGN). We can rewrite (\ref{model_1}) in matrix form as
\begin{equation}
\label{model_matrix}
\mathbf{r}_m = \sum_{i=1}^{N_t}\mathbf{S}_i\mathbf{h}_{i,m}+\mathbf{z}_m,
\end{equation}
where $\mathbf{r}_m$ and $\mathbf{z}_m$ have dimension $N\times1$, $\mathbf{h}_{i,m}$ is of dimension $L\times1$, and $\mathbf{S}_i$ is Toeplitz matrix of dimension $N\times L$ defined as

\begin{equation}
\mathbf{S}_i=
\left[\begin{array}{ccc}
s_i[0]& \cdots &s_i[1-L]\\
s_i[1]& \cdots &s_i[2-L]\\
\vdots& \vdots &\vdots\\
s_i[L-1]& \cdots &s_i[0]\\
\vdots& \vdots &\vdots\\
s_i[N-1]& \cdots &s_i[N-L]
\end{array}\right].
\end{equation}

Matched filters are used at the receivers to estimate the channel response, as illustrated in Fig. \ref{fig_egsounding}. For estimating the channel response $\mathbf{h}_{i,m}$, the matched filter in matrix form is $\mathbf{S}_i^H$. We can express the channel estimation as
\begin{equation}
\label{matched_filter}
\mathbf{h}_{i,m}^I = \mathbf{S}_i^H\mathbf{r}_m=\mathbf{S}_i^H\mathbf{S}_i\mathbf{h}_{i,m}+\sum_{v\neq i}\mathbf{S}_i^H\mathbf{S}_v\mathbf{h}_{v,m}+\mathbf{S}_i^H\mathbf{z}_m
\end{equation}
The $\mathbf{S}_i^H\mathbf{S}_i$ term in (\ref{matched_filter}) is the auto-correlation matrix for the $i^{th}$ channel sounding sequence
\begin{equation}
\mathbf{S}_i^H\mathbf{S}_i=
\left[\begin{array}{ccc}
R_i[0]& \cdots &R_i[L-1]\\
R_i[-1]& \cdots &R_i[L-2]\\
\vdots& \vdots &\vdots\\
R_i[-L+1]& \cdots &R_i[0]\\
\end{array}\right]
\end{equation}
where $R_i[\tau]$ is the periodic autocorrelation function for the $i^{th}$ waveform, which is defined as \cite{Sarwate1979}
\begin{equation}
\label{Rtau}
R_i[\tau] = \sum_{n=0}^{N-\tau-1}s_i[n]s_i^{\ast}[n+\tau] + \sum_{n=N-\tau}^{N-1}s_i[n]s_i^{\ast}[n+\tau-N].
\end{equation}
The $\mathbf{S}_i^H\mathbf{S}_v$ term in (\ref{matched_filter}) is the cross-correlation matrix between the $i^{th}$ and the $v^{th}$ channel sounding waveforms, which can be written as
\begin{equation}
\mathbf{S}_i^H\mathbf{S}_v=
\left[\begin{array}{ccc}
C_{iv}[0]& \cdots &C_{iv}[L-1]\\
C_{iv}[-1]& \cdots &C_{iv}[L-2]\\
\vdots& \vdots &\vdots\\
C_{iv}[-L+1]& \cdots &C_{iv}[0]\\
\end{array}\right],
\end{equation}
where $C_{iv}[\tau]$ is the periodic cross-correlation function between the $i_{th}$ and $v_{th}$ sequences, which is defined as \cite{Sarwate1979}
\begin{equation}
\label{Rtau}
C_{iv}[\tau] = \sum_{n=0}^{N-\tau-1}s_i[n]s_v^{\ast}[n+\tau] + \sum_{n=N-\tau}^{N-1}s_i[n]s_v^{\ast}[n+\tau-N].
\end{equation}
\subsection{With fractional offset}
Next, we consider the scenario with fractional offset. Due to the existence of a fractional offset, pulse shaping filter has to be taken into account here. Let $g(t)$ denote the pulse shaping filter defined on the interval $[-MT, MT]$, the transmitted continuous-time waveform at the $i^{th}$ antenna $s_i(t)$ can be expressed as
\begin{equation}
\label{model_12}
s_i(t)= \sum_{n=-\infty}^{\infty}s_i[n]g(t-nT).
\end{equation}
At the receiver, the $n^{th}$ sample at the receive antenna $m$ can be expressed as
\begin{eqnarray}
\label{model_122}
r_m^{F}[n] &=& r_m^{F}(nT + \mu_{i,m}T)\\\notag
&=&\sum_{i=1}^{N_t}\sum_{l=0}^{L-1}h_{i,m}[l]s_i(nT+\mu_{i,m}T-lT)+z_m[n]\\\notag
&=& \sum_{i=1}^{N_t}\sum_{l=0}^{L-1}h_{i,m}[l]\sum_{k=-\infty}^{\infty}s_i[k]g(nT+\mu_{i,m}T-lT-kT)+z_m[n] ,\,\,n=0,\dots,N-1.
\end{eqnarray}
Let $y =n-k$. Since $g(t)$ is only defined on the interval $[-MT, MT]$ and $\mu_{i,m} \in (0, \frac{1}{2}]$, we can obtain
\begin{equation}
-MT \leq yT+\mu_{i,m}T-lT \leq MT.
\end{equation}
Since $0 \leq l \leq L-1$, we can further obtain
\begin{equation}
-M \leq y\leq M+L-2.
\end{equation}
Plugging $y =n-k$ into Eq. (\ref{model_122}), we can rewrite (\ref{model_122}) as
\begin{eqnarray}
\label{model_re2}
r_m^{F}[n] = \sum_{i=1}^{N_t}\sum_{l=0}^{L-1}\sum_{y=-M}^{M+L-2}s_i[n-y]g(yT+\mu_{i,m}T-lT)h_{i,m}[l]+z_m[n] ,\,\,n=0,\dots,N-1.
\end{eqnarray}
The Eq. (\ref{model_re2}) can be rewritten in matrix form as
\begin{equation}
\label{model_matrix2}
\mathbf{r}_m^{F} = \sum_{i=1}^{N_t}\mathbf{S}_i^{F}\mathbf{G}_{\mu_{i,m}}\mathbf{h}_{i,m}+\mathbf{z}_m,
\end{equation}
where $\mathbf{r}_m^{F} = \left[r_m^F[0], r_m^F[1], \cdots, r_m^F[N-1]\right]^T $ has dimension $N\times1$, $\mathbf{S}_i^{f}$ is Toeplitz matrix of dimension $N\times (2M+L-1)$ defined as
\begin{equation}
\mathbf{S}_i^{F}=
\left[\begin{array}{ccccc}
s_i[M]& \cdots & s_i[0] &\cdots &s_i[-M-L+2]\\
s_i[M+1]& \cdots & s_i[1] &\cdots &s_i[-M-L+3]\\
\vdots& \vdots & \vdots &\vdots &\vdots\\
s_i[M+N-1]& \cdots & s_i[N-1] &\cdots & s_i[-M-L+N+1]
\end{array}\right],
\end{equation}
and $\mathbf{G}(\mu_{i,m})$ is Toeplitz matrix of dimension $(2M+L-1)\times L$ defined as
\begin{equation}
\mathbf{G}(\mu_{i,m})=
\left[\begin{array}{ccc}
g(-MT + \mu_{i,m}T)& \cdots &g((-M-L+1)T + \mu_{i,m}T)\\
\vdots& \vdots &\vdots\\
g(-T + \mu_{i,m}T)& \cdots &g_i(-LT + \mu_{i,m}T)\\
g(\mu_{i,m}T)& \cdots &g_i(-(L-1)T + \mu_{i,m}T)\\
g(T + \mu_{i,m}T)& \cdots &g_i(-(L-2)T + \mu_{i,m}T)\\
\vdots& \vdots &\vdots\\
g((M+L-2)T + \mu_{i,m}T)& \cdots &g_i((M-1)T + \mu_{i,m}T)
\end{array}\right].
\end{equation}
By left multiplying received vector $\mathbf{r}_m^{F}$ with the matched filter $(\mathbf{S}_i^{F})^H$, the output can be expressed as
\begin{equation}
\label{matched_filter2}
\mathbf{h}_{i,m}^{F} = (\mathbf{S}_i^{F})^H\mathbf{r}_m^f=(\mathbf{S}_i^{F})^H\mathbf{S}_i^{F}\mathbf{G}(\mu_{i,m})\mathbf{h}_{i,m}+\sum_{v\neq i}(\mathbf{S}_i^{F})^H\mathbf{S}_v^{F}\mathbf{G}(\mu_{i,m})\mathbf{h}_{v,m}+(\mathbf{S}_i^{F})^H\mathbf{z}_m,
\end{equation}
where $(\mathbf{S}_i^{F})^H\mathbf{S}_i^{F}$ and $(\mathbf{S}_i^{F})^H\mathbf{S}_v^{F}$ are of dimension $(2M+L-1)\times (2M+L-1)$ expressed as
\begin{equation}
(\mathbf{S}_i^{F})^H\mathbf{S}_i^{F}=
\left[\begin{array}{ccc}
R_i[0]& \cdots &R_i[2M+L-2]\\
R_i[-1]& \cdots &R_i[2M+L-3]\\
\vdots& \vdots &\vdots\\
R_i[-2M-L+2]& \cdots &R_i[0]\\
\end{array}\right]
\end{equation}
\begin{equation}
(\mathbf{S}_i^{F})^H\mathbf{S}_v^{F}=
\left[\begin{array}{ccc}
C_{iv}[0]& \cdots &C_{iv}[2M+L-2]\\
C_{iv}[-1]& \cdots &C_{iv}[2M+L-3]\\
\vdots& \vdots &\vdots\\
C_{iv}[-2M-L+2]& \cdots &C_{iv}[0]\\
\end{array}\right]
\end{equation}

\section{Objective of Sounding Waveforms Design}
In this section, we show the conditions that the sounding waveform must meet in order to independently resolve all of the spatial channel responses.

Firstly, with only integer offset, the estimated channel response in (\ref{matched_filter}) consists of  both the desired signal term $\mathbf{S}_i^H\mathbf{S}_i\mathbf{h}_{i,m}$ and the error term $\sum_{v\neq i}\mathbf{S}_i^H\mathbf{S}_v\mathbf{h}_{v,m}+\mathbf{S}_i^H\mathbf{z}$. The ideal situation would be $\mathbf{S}_i^T\mathbf{S}_i\mathbf{h}_{i,m} = \mathbf{h}_{i,m}$ and $\sum_{v\neq i}\mathbf{S}_i^H\mathbf{S}_v\mathbf{h}_{v,m}=\mathbf{0}$. Hence, the objective is to make the waveforms satisfy
\begin{equation}
\mathbf{S}_i^H\mathbf{S}_v =
\begin{cases}
\mathbf{I}_{L\times L},\quad \text{if} \quad i = v\\
\mathbf{0}_{L\times L}, \quad \text{if} \quad i \neq v
\end{cases}
\end{equation}
which is equivalent to the condition that
\begin{equation}
\label{rcond}
R_i[\tau] =
\begin{cases}
1, \quad \text{if} \quad \tau=0\\
0, \quad \text{if} \quad 0 < |\tau|<L
\end{cases}
\end{equation}
\begin{equation}
\label{cijcond}
C_{iv}[\tau] = 0, \quad \text{if} \quad -L<\tau<L
\end{equation}
It is worthwhile to point out that the criteria for an asynchronous MU-MIMO system in (\ref{rcond}) and (\ref{cijcond}) are stricter than those for a synchronous MU-MIMO system. In the synchronous MU-MIMO system, since all the transmitters and receivers are synchronized, it is unnecessary to model the effects of clock mismatch delays as part of the channel response. Hence, the overall length of channel response $L$ can be set as small as $L=\underset{i,m}{\max}\{L_{i,m}\}$. However, in the asynchronous MU-MIMO system, we have to take the clock mismatch delays into account, and the minimum
$L$ can be set as $L=\underset{i,m}{\max}\{L_{i,m}+d_{i,m}\}$.

Secondly, with fractional offset, the interference from other channels $\sum_{v\neq i}(\mathbf{S}_i^{F})^H\mathbf{S}_v^{F}\mathbf{G}(\mu_{i,m})\mathbf{h}_{v,m}$ in (\ref{matched_filter2}) can be eliminated by making $(\mathbf{S}_i^{F})^H\mathbf{S}_v^{F} = \mathbf{0}_{(2M+L-1)\times (2M+L-1)}, \, i \neq v$. However, $\mathbf{h}_{i,m}$ can not be obtained from the signal term $(\mathbf{S}_i^{F})^H\mathbf{S}_i^{F}\mathbf{G}(\mu_{i,m})\mathbf{h}_{i,m}$ since $\mu_{i,m}$ is unknown. Let the waveform satisfy
\begin{equation}
(\mathbf{S}_i^F)^H\mathbf{S}_v^{F} =
\begin{cases}
\mathbf{I}_{(2M+L-1)\times (2M+L-1)}  ,\quad \text{if} \quad i = v\\
\mathbf{0}_{(2M+L-1)\times (2M+L-1)}, \quad \text{if} \quad i \neq v
\end{cases}
\end{equation}
which is equivalent to the condition that
\begin{equation}
\label{rcond2}
R_i[\tau] =
\begin{cases}
1, \quad \text{if} \quad \tau=0\\
0, \quad \text{if} \quad |\tau| < 2M+L-1,\, \text{and} \,\, \tau \neq 0
\end{cases}
\end{equation}
\begin{equation}
\label{cijcond2}
C_{iv}[\tau] = 0, \quad \text{if} \quad |\tau| <2M+L-1
\end{equation}

For comparison purposes, we list various scenarios and the constraints they impose on the optimal sounding waveforms in Table \ref{table_constraints}.


\begin{table}[!t]
\renewcommand{\arraystretch}{1.3}
\caption{Constraints on optimal sounding waveforms for various scenarios}
\label{table_constraints}
\centering
  \begin{tabular}{p{4cm}|c|c}
  \hline
    Configuration & Auto correlation requirements & Cross correlation requirements\\
    \hline
    \hline
    SU-SISO \cite{Spasojevic2001} & $R_1[\tau] = \begin{cases} 1, \,\,\tau=0\\0, \,\,|\tau|<L_{1,1} \end{cases}$ & N/A\\
    \hline
    SU-MIMO \quad \quad or Synchronous MU-SISO or Synchronous MU-MIMO \cite{Deng2004} & $R_i[\tau] = \begin{cases} 1, \,\,\tau=0\\0, \,\,|\tau|<\underset{i,m}{\max}\{L_{i,m}\}\end{cases}$ & $C_{iv}[\tau] = 0, \,\,|\tau|<\underset{i,m}{\max}\{L_{i,m}\}$\\
    \hline
    Asynchronous MU-SISO or Asynchronous MU-MIMO with only integer offset& $R_i[\tau)] = \begin{cases} 1, \,\,\tau=0\\0, \,\,|\tau|<\underset{i,m}{\max}\{L_{i,m}+d_{i,m}\} , \end{cases}$ & $C_{iv}[\tau] = 0, \,\,|\tau|<\underset{i,m}{\max}\{L_{i,m}+d_{i,m}\}$\\
    \hline
    Asynchronous MU-SISO or Asynchronous MU-MIMO with fractional offset& $R_i[\tau] = \begin{cases} 1, \,\,\tau=0\\0, \,\, |\tau| < \underset{i,m}{\max}\{L_{i,m}+d_{i,m}\} + 2M -1  \end{cases}$ & $C_{iv}[\tau] = 0,\,\,|\tau|<\underset{i,m}{\max}\{L_{i,m}+d_{i,m}\}+ 2M -1$\\
    \hline
  \end{tabular}
\end{table}
Additionally, since a high peak-to-average power ratio (PAPR) of a transmitted waveform dramatically reduces the efficiency of the communication system, we desire low PAPR waveforms. The PAPR of the waveform $\mathbf{s_i}$ is defined as
\begin{eqnarray}
\text{PAPR}(\mathbf{s_i})=\frac{\max|s_i[n]|^2}{\sum_{n=1}^N |s_i[n]|^2/N}.
\end{eqnarray}

\section{Sounding Waveform Design}
To achieve minimum mean square error (MSE) in the channel sounding for asynchronous MU-MIMO systems, the waveforms transmitted from the multiple antennas must have impulse-like auto correlation function
and zero cross correlation as indicated in (\ref{rcond}), (\ref{cijcond}), (\ref{rcond2}) and (\ref{cijcond2}). The zero cross correlation requirement is due to the possibility that the mismatch delay, $d_{i,m}$, may be very large, i.e. the maximum of $L_{i,m}+d_{i,m}$ is essentially unbounded. In \cite{Fragouli2003, Deng2004} a constant-magnitude polyphase channel sounding sequence was proposed which has minimum PAPR but does not satisfy the minimum MSE condition. The advantages of using chirp waveforms, including the low PAPR property, have been recognized in semi-sequential MIMO sounders \cite{Salous2002,Salous2005,Salous2010} and other channel sounders \cite{Uryadov1995,Salous1998,Davydenko2008,Mar2009,Jong2010,Kun2010,Zhang2011}. In this paper, we construct the chirp waveform set to satisfy the minimum MSE conditions for asynchronous MU-MIMO channel sounding. We design the channel sounding chirp waveforms as
\begin{eqnarray}
s^{(p)}[n] = \frac{1}{\sqrt{N}}\exp\left\{j2\pi\frac{p}{N}(n+1)(n+2)\right\},\quad n=0,\dots,N-1,
\end{eqnarray}
where $j = \sqrt{-1}$. We restrict both the $p$ and $N$ to be powers of 2 and $N > 2p$, thus, the ratio $N/(2p)$ is always even. The auto-correlation $R^{(p)}[\tau]$ and cross-correlation $C^{(p,q)}[\tau]$ are proved in the appendix to be
\begin{equation}
\label{Rindex}
R^{(p)}[\tau] =
\begin{cases}
1, \qquad \text{if} \qquad \tau = w\frac{N}{p}\\
-1, \qquad \text{if} \qquad \tau = (2w+1)\frac{N}{2p}\\
0, \qquad \text{otherwise}
\end{cases}
\end{equation}
\begin{equation}
w = 0,\dots,(p-1)
\end{equation}
\begin{eqnarray}
C^{(p,q)}[\tau] = 0, \, \tau=0,1,\dots,N-1
\end{eqnarray}
To satisfy (\ref{rcond}), we require
\begin{equation}
\label{descr}
N>2p_{\max}L;
\end{equation}
To satisfy (\ref{rcond2}), we require
\begin{equation}
\label{descr}
N>2p_{\max}(2M+L-1),
\end{equation}
where $p_{\max}$ is the largest value of $p$ for the set of MU-MIMO channel sounding waveforms.

With only integer offset, the estimation of channel response in (\ref{matched_filter}) becomes
\begin{equation}
\label{mse_1}
\mathbf{h}_{i,m}^I = \mathbf{h}_{i,m}+\mathbf{S}_i^H\mathbf{z}_m
\end{equation}
The mean square error at the receive antenna $m$ can be obtained as
\begin{equation}
\label{mse_2}
\text{MSE}_m= E\bigg\{\|\mathbf{h}_{i,m} - \mathbf{h}_{i,m}^I\|^2\bigg\} = E\bigg\{\|\mathbf{S}_i^H\mathbf{z}_m\|^2\bigg\}
\end{equation}
We assume that AWGN $\mathbf{z}_m$ has auto-correlation matrix $R_{z_m} = E\{\mathbf{z}_m\mathbf{z}_m^H\}=2\sigma_m^2\mathbf{I}_N$. Hence the noise term $\mathbf{S}_i^H\mathbf{z}$ has auto-correlation matrix $R_z = E\{\mathbf{S}_i^H\mathbf{z}\mathbf{z}^H\mathbf{S}_i\}=2\sigma^2\mathbf{I}_L$. The $\text{MSE}_m$ in (\ref{mse_2}) can be further derived as
\begin{equation}
\label{mse_3}
\text{MSE}_m= E\bigg\{\|\mathbf{S}_i^H\mathbf{z}_m\|^2\bigg\}=E\bigg\{\mathbf{z}_m^H\mathbf{S}_i\mathbf{S}_i^H\mathbf{z}_m\bigg\} = 2L\sigma_m^2
\end{equation}
We have derived the Cramer-Rao bound (see Appendix) as
\begin{equation}
\label{crb1}
\text{CRB}_m = 2L\sigma_m^2.
\end{equation}
The results from (\ref{mse_3}) shows that the MSE of our system meets the CRB.

With fractional offset, the output of the matched filter in (\ref{matched_filter2}) becomes
\begin{equation}
\mathbf{h}_{i,m}^{F} = \mathbf{G}(\mu_{i,m})\mathbf{h}_{i,m}+(\mathbf{S}_i^{F})^H\mathbf{z}_m.
\end{equation}
We can see that the channel response vector $\mathbf{h}_{i,m}$ is multiplied by the pulse shaping matrix $\mathbf{G}(\mu_{i,m})$, which is a function of the fractional offset $\mu_{i,m}$. The noise term $(\mathbf{S}_i^F)^H\mathbf{z}_m$ has auto-correlation matrix $R_z = E\{(\mathbf{S}_i^{F})^H\mathbf{z}_m\mathbf{z}_m^H\mathbf{S}_i^{F}\}=2\sigma^2\mathbf{I}_{2M+L-1}$. Given the observations $\mathbf{h}_{i,m}^{F}$, which is a $(2M+L-1)$-length vector, we need to jointly estimate $\mu_{i,m}$ and $L$-length channel response vector $\mathbf{h}_{i,m}$. The estimation of $\mu_{i,m}$ and  $\mathbf{h}_{i,m}$ is given by
\begin{equation}
\label{eq_argminmu}
\{\hat{\mu}_{i,m}, \hat{\mathbf{h}}_{i,m}\} = \arg \underset{\mu, \mathbf{h}}{\min} ||\mathbf{h}_{i,m}^{F} - \mathbf{G}(\mu)\mathbf{h}||_2^2.
\end{equation}
The iterative solution of Eq. (\ref{eq_argminmu}) can be expressed as
\begin{equation}
\label{eq_muest}
\hat{\mu}_{i,m}^{(n+1)} = \arg \underset{\mu}{\min} ||\mathbf{h}_{i,m}^{F} - \mathbf{G}(\mu)\hat{\mathbf{h}}_{i,m}^{(n)}||_2^2,
\end{equation}
\begin{equation}
\label{eq_hest}
\hat{\mathbf{h}}_{i,m}^{(n+1)} = \left(\mathbf{G}^H\left(\hat{\mu}_{i,m}^{(n+1)}\right)\mathbf{G}\left(\hat{\mu}_{i,m}^{(n+1)}\right)\right)^{-1}\mathbf{G}^H\left(\hat{\mu}_{i,m}^{(n+1)}\right)\mathbf{h}_{i,m}^{F}.
\end{equation}
Since $g(t)$ is a differentiable function, Newton's method can be employed to solve the minimization problem in Eq. (\ref{eq_muest}).

\subsection{Channel Estimate Averaging}
The above analysis only considers at most the $(2M+L-1)$-length
matched filter output. As is apparent from (\ref{Rindex}), each matched filter output contains $2p$ copies of the channel impulse response estimate.  Based on this, a natural question arises: can the MSE of the channel estimation be decreased by averaging these $2p$ outputs?  Next we will show that averaging the multiple segments of the matched filter output {\it can not} refine the above results and
each branch has the same channel sounding performance.  It is possible to drive down the MSE by averaging the outputs from multiple sequences as long as the channel is static across the soundings.  We only show that inter-symbol averaging can not decrease the MSE.

If we assume the channel is static for at least $N$-length matched filter output, we can write the $N$-length output as
\begin{eqnarray}
\label{Noutput}
\mathbf{h}_{i,m}^{L} &=& \left[\begin{array}{ccc}
s_i[M]& \cdots &s_i[M+1-N]\\
s_i[M+1]& \cdots &s_i[M+2-N]\\
\vdots& \vdots &\vdots\\
s_i[M+N-1]& \cdots &s_i[M]
\end{array}\right]^H\mathbf{r}_{m}^{F}\\\notag
&=&  \left[\begin{array}{ccc}
R_i[0]& \cdots &R_i[2M+L-2]\\
R_i[-1]& \cdots &R_i[2M+L-3]\\
\vdots& \vdots &\vdots\\
R_i[-N+1]& \cdots &R_i[2M+L-N-1]\\
\end{array}\right]\mathbf{G}(\mu_{i,m})\mathbf{h}_{i,m}\\\notag
&&+\mathbf{\tilde{z}}_m
\end{eqnarray}
where $\mathbf{\tilde{z}}_m$ denotes the noise part of the $N$-length output.
\begin{eqnarray}
\label{Nnoise}
\mathbf{\tilde{z}}_m
= \left[\begin{array}{ccc}
s_i[M]& \cdots &s_i[M+1-N]\\
s_i[M+1]& \cdots &s_i[M+2-N]\\
\vdots& \vdots &\vdots\\
s_i[M+N-1]& \cdots &s_i[M]
\end{array}\right]^H\mathbf{z}_m
\end{eqnarray}
By substituting (\ref{Rindex}) into (\ref{Noutput}), the $N$-length output can be further written as
\begin{eqnarray}
\label{multiNoutput}
\mathbf{h}_{i,m}^{l} &=& [
\mathbf{h}_{i,m}^H\mathbf{G}(\mu_{i,m})^H, \mathbf{0}_{1\times (\frac{N}{2p}-2M-L+1)}, -\mathbf{h}_{i,m}^H\mathbf{G}(\mu_{i,m})^H, \mathbf{0}_{1\times (\frac{N}{2p}-2M-L+1)},\dots,\\\notag
&&\mathbf{h}_{i,m}^H\mathbf{G}(\mu_{i,m})^H, \mathbf{0}_{1\times (\frac{N}{2p}-2M-L+1)},
-\mathbf{h}_{i,m}^H\mathbf{G}(\mu_{i,m})^H, \mathbf{0}_{1\times (\frac{N}{2p}-2M-L+1)}]^H \\\notag
&&+ \mathbf{\tilde{z}}_m
\end{eqnarray}
We can see from (\ref{multiNoutput}) that the $N$-Length output $\mathbf{h}_{i,m}^{L}$ consists of $2p$ replicas of the channel response $\mathbf{G}(\mu_{i,m})\mathbf{h}_{i,m}$ plus the filtered noise $\mathbf{\tilde{z}}_m$. If the noise $\mathbf{\tilde{z}}_m$ were white, we could increase the SNR by a factor of $2p$ by averaging the $2p$ segments of the matched filter output. However, the noise $\mathbf{\tilde{z}}_m$ has the auto-correlation matrix
\begin{eqnarray}
\label{noiseauto}
R_{\tilde{z}_m} &=& E\{\mathbf{\tilde{z}}_m\mathbf{\tilde{z}}_m^H\}\\\notag
&=&2\sigma_m^2\left[\begin{array}{ccc}
R_i(0)& \cdots &R_i(N-1)\\
R_i(-1)& \cdots &R_i(N-2)\\
\vdots& \vdots &\vdots\\
R_i(-N+1)& \cdots &R_i(0)\\
\end{array}\right],
\end{eqnarray}
which is not a multiple of the identity matrix. In addition, according to (\ref{Rindex}) and (\ref{noiseauto}), the noise $\mathbf{\tilde{z}}_m$ has auto-correlation function values $R[N/2p] = -2\sigma_m^2$
and $R[N/p] = 2\sigma_m^2$.
Therefore, averaging the $2p$ output segments with the distance $N/2p$ can not increase the SNR or improve the MSE in (\ref{mse_3}).
The above analysis indicates that the performance of channel sounding is independent of the parameter $p$ adopted by the waveform.


\subsection{Capacity Analysis}
Let $\tilde{h}_{i,m}(f)$ denote the channel frequency response at frequency $f$ of the synchronous MU-MIMO system and $\breve{h}_{i,m}(f)$ denote the channel frequency response at frequency $f$ of the asynchronous MU-MIMO system. The equations $\tilde{h}_{i,m}(f)$ and $\breve{h}_{i,m}(f)$ should satisfy
\begin{equation}
\label{hhff}
\breve{h}_{i,m}(f) = e^{-j2\pi f\zeta_{i,m}}\tilde{h}_{i,m}(f).
\end{equation}
The matrix form of the channel frequency response at frequency $f$ for the synchronous MU-MIMO and asynchronous MU-MIMO can be written as
\begin{equation}
\label{channelf}
\mathbf{\tilde{H}}(f)= \left[\begin{array}{cccc}
\tilde{h}_{1,1}(f)&\tilde{h}_{2,1}(f)& \cdots &\tilde{h}_{N_t,1}(f)\\
\tilde{h}_{1,2}(f)& \ddots &&\vdots\\
\vdots& &\ddots&\vdots\\
\tilde{h}_{1,N_r}(f)& \cdots & \cdots &\tilde{h}_{N_t,N_r}(f)\\
\end{array}\right]
\end{equation}
\begin{equation}
\label{channelf}
\mathbf{\breve{H}}(f)= \left[\begin{array}{cccc}
\breve{h}_{1,1}(f)&\breve{h}_{2,1}(f)& \cdots &\breve{h}_{N_t,1}(f)\\
\breve{h}_{1,2}(f)& \ddots &&\vdots\\
\vdots& &\ddots&\vdots\\
\breve{h}_{1,N_r}(f)& \cdots & \cdots &\breve{h}_{N_t,N_r}(f)\\
\end{array}\right].
\end{equation}
The capacity of the synchronous MU-MIMO channel and synchronous MU-MIMO channel are given by
\begin{equation}
\label{capacity1}
C_{syn} = \frac{1}{B}\int_B\log_2\det\left(\mathbf{I}+\frac{\rho}{N_t}\mathbf{\tilde{H}}^H(f)\mathbf{\tilde{H}}(f)\right)df
\end{equation}
\begin{equation}
\label{capacity2}
C_{asyn} = \frac{1}{B}\int_B\log_2\det\left(\mathbf{I}+\frac{\rho}{N_t}\mathbf{\breve{H}}^H(f)\mathbf{\breve{H}}(f)\right)df,
\end{equation}
where $\rho$ denotes the signal-to-noise ratio (SNR) and B is the signal bandwidth.

If the $M_t$ transmitters are synchronized or they share a single LO, the clock mismatch delays should satisfy
\begin{equation}
\kappa_m = \zeta_{1,m} = \zeta_{2,m} = \cdots = \zeta_{N_t,m}, \,\, m = 1,2,\dots,N_r.
\end{equation}
From (\ref{hhff}), we can have
\begin{eqnarray}
\mathbf{\breve{H}}(f) &=& \left[\begin{array}{cccc}
e^{-j2\pi f\kappa_1}&0& \cdots &0 \\
0& e^{-j2\pi f\kappa_2} &&\vdots\\
\vdots& &\ddots&\vdots\\
0& \cdots & \cdots &e^{-j2\pi f\kappa_{N_r}}\\
\end{array}\right]\mathbf{\tilde{H}}(f) \triangleq \mathbf{A}\mathbf{\tilde{H}}(f).
\end{eqnarray}
Thus, we can obtain
\begin{eqnarray}
\mathbf{\breve{H}}^H(f)\mathbf{\breve{H}}(f) &=& \mathbf{\tilde{H}}^H(f)\mathbf{A}^H\mathbf{A}\mathbf{\tilde{H}}(f)\\\notag
&=& \mathbf{\tilde{H}}^H(f)\mathbf{\tilde{H}}(f).
\end{eqnarray}

If the $M_r$ receivers are synchronized or they share a single LO, the clock mismatch delays should satisfy
\begin{equation}
\iota_i = \zeta_{i,1} = \zeta_{i,2} = \cdots = \zeta_{i,N_r}, \,\, i = 1,2,\dots,N_t.
\end{equation}
From (\ref{hhff}), we have
\begin{eqnarray}
\mathbf{\breve{H}}(f) &=& \mathbf{\tilde{H}}(f)\left[\begin{array}{cccc}
e^{-j2\pi f\iota_1}&0& \cdots &0 \\
0& e^{-j2\pi f\iota_2} &&\vdots\\
\vdots& &\ddots&\vdots\\
0& \cdots & \cdots &e^{-j2\pi f\iota_{N_t}}\\
\end{array}\right] \triangleq \mathbf{\tilde{H}}(f)\mathbf{B}.
\end{eqnarray}
Thus, we obtain
\begin{eqnarray}
\mathbf{\breve{H}}^H(f)\mathbf{\breve{H}}(f) = \mathbf{B}^H\mathbf{\tilde{H}}^H(f)\mathbf{\tilde{H}}(f)\mathbf{B}.
\end{eqnarray}
Since $\mathbf{B}$ is a unitary matrix, we have
\begin{eqnarray}
\det\left(\mathbf{I}+\frac{\rho}{N_t}\mathbf{\breve{H}}^H(f)\mathbf{\breve{H}}(f)\right) &=&
\det\left(\mathbf{I}+\frac{\rho}{N_t}
\mathbf{B}^H\mathbf{\tilde{H}}^H(f)\mathbf{\tilde{H}}(f)\mathbf{B}\right) \\\notag
&=& \det\left(\mathbf{I}+\frac{\rho}{N_t}\mathbf{\tilde{H}}^H(f)\mathbf{\tilde{H}}(f)\right).
\end{eqnarray}
From (\ref{capacity1}) and (\ref{capacity2}), we can conclude that the capacity region is the same as long as one side has a single LO. However, if both sides have multiple LOs,
\begin{eqnarray}
\det\left(\mathbf{I}+\frac{\rho}{N_t}\mathbf{\breve{H}}^H(f)\mathbf{\breve{H}}(f)\right) \neq \det\left(\mathbf{I}+\frac{\rho}{N_t}\mathbf{\tilde{H}}^H(f)\mathbf{\tilde{H}}(f)\right),
\end{eqnarray}
the capacity region is not the same. We need to synchronize the LOs at either the transmitter side or receiver side to determine the capacity.

\section{Simulation}
\begin{figure}[!t]
\centering
\includegraphics[width=10cm]{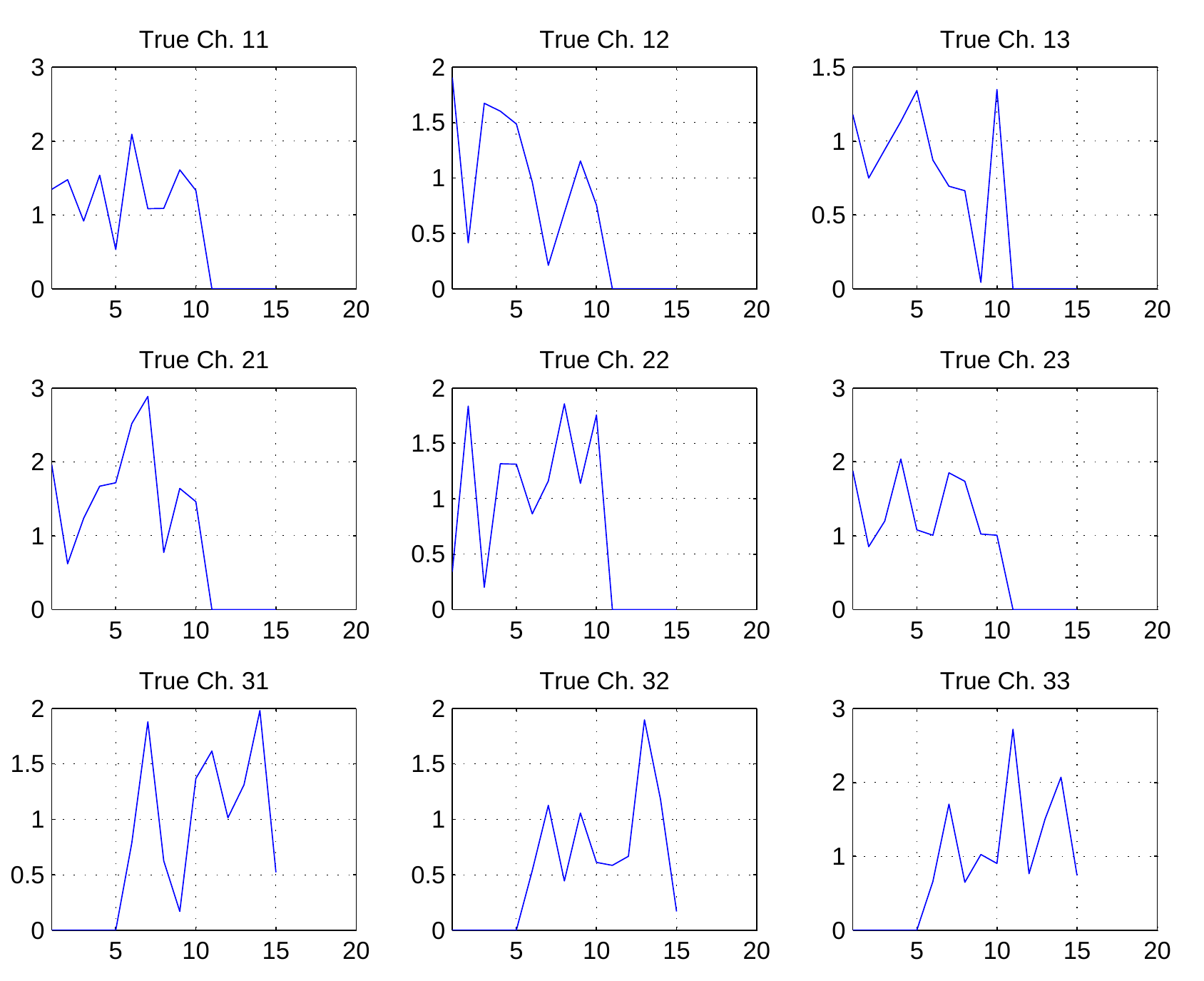}
\hfil
\caption{Multipath channels used in the simulations}
\label{fig_channelsimulation1}
\end{figure}

\begin{figure}[!t]
\centering
\includegraphics[width=3.8in]{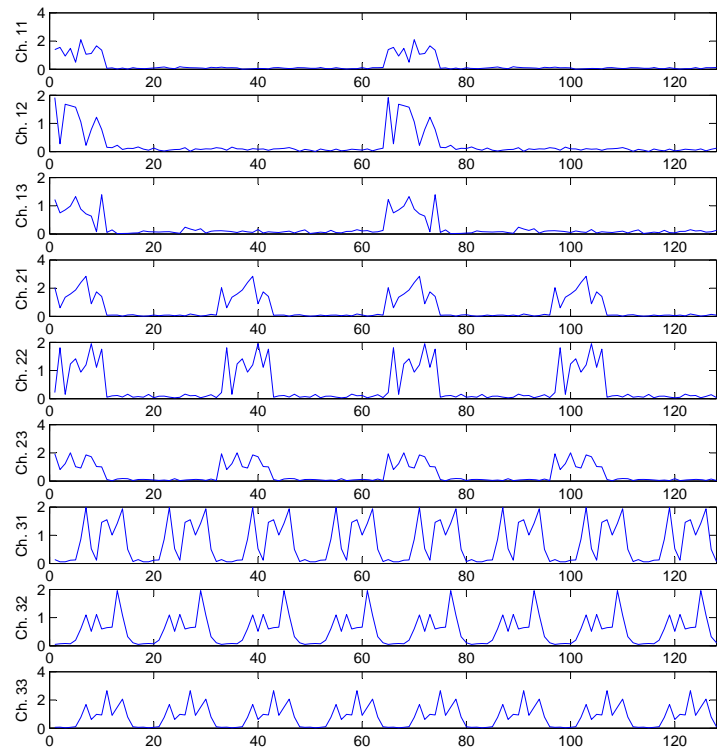}
\caption{128-length raw output of sounded channel responses}
\label{fig_channelsimulation2}
\end{figure}

\begin{figure}[!t]
\centering
\includegraphics[width=3.3in]{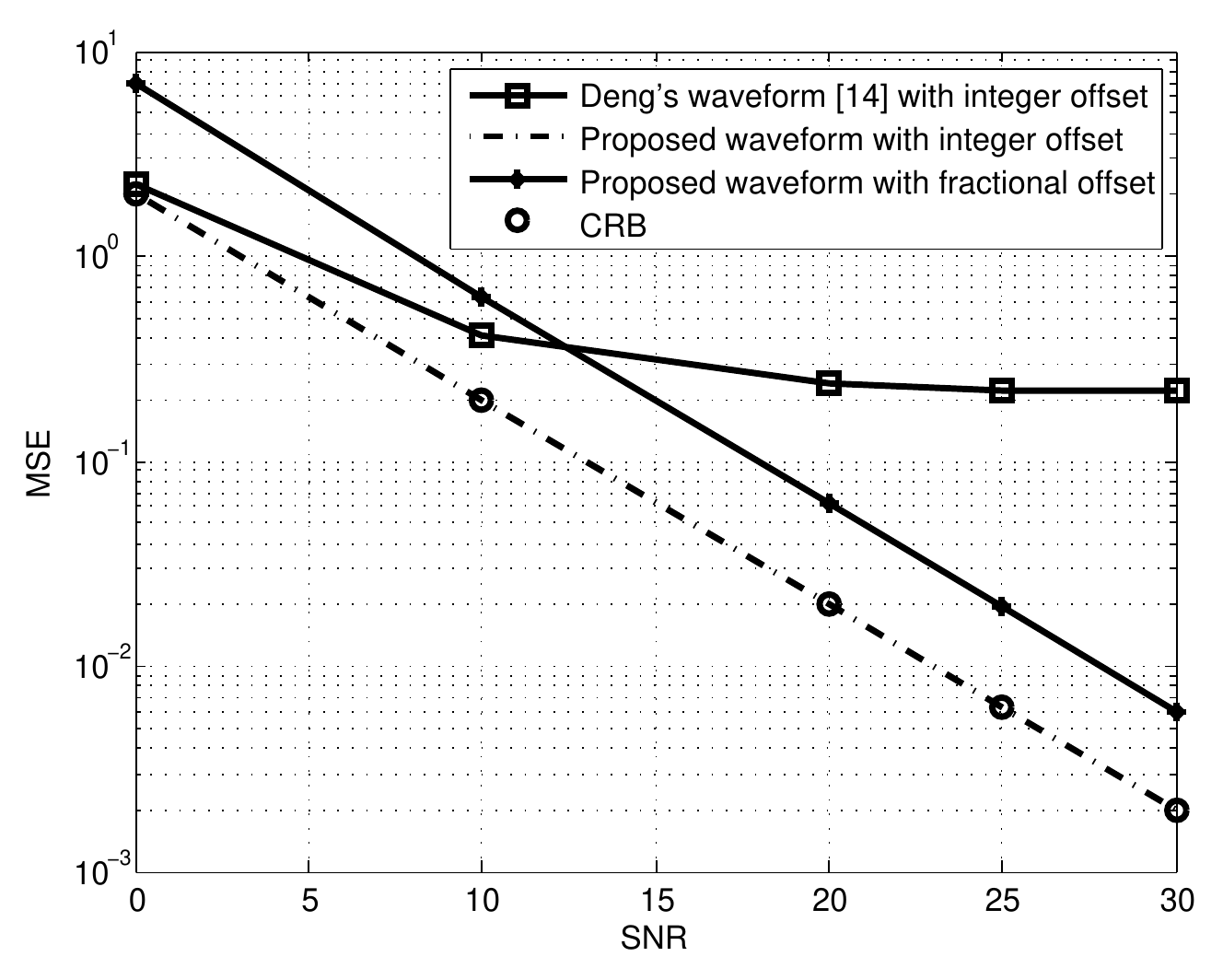}
\caption{MSE of channel sounding}
\label{fig_channelsimulation3}
\end{figure}

We set up the simulation scenario as follows: $M_t = 2, N_t = 3, M_r = 3, N_r = 3$.
Thus, we have 9 channels to be generated. We set the integer offset for the first user to be zero and the second user to be 5. All the 9 channels have 10 nonzero taps. The AWGN is assumed to have the same $\text{SNR} = 25$ dB at the 3 receive antennas in the simulation. To make sure that the sounding waveform satisfies the criteria in (\ref{descr}), we use the parameters as: $N = 128, p = 1, 2, 4$ for the three waveforms respectively. The generated channel responses are shown in Fig \ref{fig_channelsimulation1}. Fig \ref{fig_channelsimulation2} shows the magnitude of the $N$-length outputs after the
matched filter. We can see that the $N$-length outputs include $2p = 2, 4, 8$
segments of the sounded channel responses. The MSE of channel sounding (generated by 10000 Monte-Carlo runs) and CRB
are presented in Fig \ref{fig_channelsimulation3}. The result with only integer offset matched CRB very well. The MSE of the proposed waveform with fractional offset is plotted as well. We randomly generated the fractional offset in each run. The performance is degraded by the unknown pulse shaping matrix. The MSE of the waveform presented in \cite{Deng2004} is also simulated (10000 Monte-Carlo runs) and compared with the proposed waveform.
We can see that the proposed chirp waveform offers a significant MSE improvement.


\section{Conclusions}
In this paper, we derived the conditions that the sounding waveform must meet in order to independently resolve all of the spatial channel responses in the MU-MIMO system. We identified a chirp waveform that meets the constraints, then we showed that the MSE of our system can meet the CRB. Finally, we demonstrated that the capacity region of the asynchronous system is the same as the perfectly synchronous collection as long as either the transmitter side or the receiver side is synchronized. This allows for a channel sounding system where only one-sided synchronization is necessary allowing for lower-cost and lower-complexity MU-MIMO channel sounding.
\clearpage

\appendices
\section{Proof of correlation of chirp waveform}

1. Auto Correlation $R^{(p)}[\tau]$
\begin{eqnarray}
R^{(p)}[\tau] &=& \sum_{n=0}^{N-\tau-1}{s^{(p)}[n]s^{\ast(p)}[n+\tau]}\\\notag
&+& \sum_{n=N-\tau}^{N-1}{s^{(p)}[n]s^{\ast(p)}[n+\tau-N]}\\\notag
&=& \sum_{n=0}^{N-1}{s^{(p)}[n]s^{\ast(p)}[n+\tau]}\\\notag
&=& \frac{1}{N}\sum_{n=0}^{N-1}\exp\bigg\{j2\pi\frac{p}{N}(n+1)(n+2)\bigg\}\\\notag
&&\exp\bigg\{-j2\pi\frac{p}{N}(n+1+\tau)(n+2+\tau)\bigg\}\\\notag
&=& \frac{1}{N}\sum_{n=0}^{N-1}\exp\bigg\{j2\pi\frac{p}{N}(-2\tau n-\tau^2-3\tau)\bigg\}\\\notag
&=& \frac{1}{N}\exp\bigg\{j2\pi\frac{p}{N}(-\tau^2-3\tau)\bigg\}\sum_{n=0}^{N-1}\bigg(\exp\bigg\{-j2\pi\frac{2p\tau}{N}\bigg\}\bigg)^n
\end{eqnarray}
where
\begin{equation}
\sum_{n=0}^{N-1}\bigg(\exp\bigg\{-j2\pi\frac{2p\tau}{N}\bigg\}\bigg)^n =
\begin{cases}
N, \qquad \tau = k\frac{N}{2p}\\
0, \qquad \text{otherwise}
\end{cases}
\end{equation}
\begin{equation}
k = 0,\dots,(2p-1)
\end{equation}
If $\tau = kN/(2p)$, $R^{(p)}[\tau]= \exp\{jk\pi(-kN/(2p)-3)\}$. Because $kN/(2p)$ is even, $R^{(p)}[\tau]= 1$ if $k$ is even and $R^{(p)}[\tau]= -1$ if $k$ is odd. Therefore,
\begin{equation}
R^{(p)}[\tau] =
\begin{cases}
1, \qquad \text{if} \qquad \tau = w\frac{N}{p}\\
-1, \qquad \text{if} \qquad \tau = (2w+1)\frac{N}{2p}\\
0, \qquad \text{otherwise}
\end{cases}
\end{equation}
\begin{equation}\notag
w = 0,\dots,(p-1)
\end{equation}

2. Cross Correlation $C^{(p,q)}[\tau]$
\begin{eqnarray}
C^{(p,q)}[\tau] &=& \sum_{n=0}^{N-\tau-1}{s^{(p)}[n])s^{\ast(q)}[n+\tau]} + \sum_{n=N-\tau}^{N-1}{s^{(p)}[n]s^{\ast(q)}[n+\tau-N]}\\\notag
&=& \sum_{n=0}^{N-1}{s^{(p)}[n]s^{\ast(q)}[n+\tau]}\\\notag
&=& \frac{1}{N}\sum_{n=0}^{N-1}\exp\bigg\{j2\pi\frac{p}{N}(n+1)(n+2)\bigg\}\\\notag
&&\exp\bigg\{-j2\pi\frac{q}{N}(n+1+\tau)(n+2+\tau)\bigg\}\\\notag
&=& \frac{1}{N}\sum_{n=0}^{N-1}\exp\bigg\{j2\pi\frac{1}{N}[(p-q)n^2\\\notag
&&-(3p-3q-2q\tau)n+(2p-2q-3q\tau-q\tau^2)]\bigg\}
\end{eqnarray}

\begin{eqnarray}
|C^{(p,q)}[\tau]|^2 &=& C^{(p,q)}[\tau]C^{(p,q)\ast}[\tau]\\\notag
&=& \frac{1}{N}\sum_{n=0}^{N-1}\sum_{m=0}^{N-1}\exp\bigg\{j2\pi\frac{1}{N}[(p-q)(n^2-m^2)\\\notag
&&-(3p-3q-2q\tau)(n-m)]\bigg\}\\\notag
&=& \frac{1}{N}\sum_{n=0}^{N-1}\sum_{m=0}^{N-1}\exp\bigg\{j2\pi\frac{1}{N}(n-m)\bigg\}\\\notag
&&\exp\bigg\{j2\pi\frac{1}{N}[(p-q)(n+m)\\\notag
&&-(3p-3q-2q\tau)]\bigg\}
\end{eqnarray}
Let $\delta = n-m$
\begin{eqnarray}\notag
|C^{(p,q)}[\tau]|^2 &=& \frac{1}{N}\sum_{n=0}^{N-1}\sum_{\delta=0}^{N-1}\exp\bigg\{j2\pi\frac{1}{N}\delta\bigg\}\\\notag
&&\exp\bigg\{j2\pi\frac{1}{N}[(p-q)(2n-\delta)\\\notag
&&-(3p-3q-2q\tau)]\bigg\}\\\notag
&=& \frac{1}{N}\sum_{\delta=0}^{N-1}\exp\bigg\{j2\pi\frac{1}{N}[-(p-q)\delta^2\\\notag
&&+(3p-3q-2q\tau)\delta]\bigg\}\\\notag
&&\sum_{n=0}^{N-1}\exp\bigg\{j2\pi\frac{2n(p-q)}{N}\delta\bigg\}\\\notag
&=& \frac{1}{N}\sum_{\delta=0}^{N-1}\exp\bigg\{j2\pi\frac{1}{N}[-(p-q)\delta^2\\\notag
&&+(3p-3q-2q\tau)\delta]\bigg\}\\\notag
&&\sum_{n=0}^{N-1}\bigg(\exp\bigg\{j2\pi\frac{2(p-q)\delta}{N}\bigg\}\bigg)^n\notag
\end{eqnarray}
where
\begin{equation}
\sum_{n=0}^{N-1}\bigg(\exp\bigg\{j2\pi\frac{2(p-q)\delta}{N}\bigg\}\bigg)^n=
\begin{cases}
N, \qquad \delta = u\frac{N}{2q}\\
0, \qquad \text{otherwise}
\end{cases}
\end{equation}
\begin{equation}
u = 0,\dots,(2q-1)
\end{equation}
Thus,
\begin{eqnarray}
|C^{(p,q)}[\tau]|^2 &=& \frac{1}{N}\sum_{u=0}^{2q-1}\exp\bigg\{j2\pi\frac{1}{N}\bigg[-(p-q)u^2\frac{N^2}{4q^2}\\\notag
&&+(3p-3q-2q\tau)u\frac{N}{2q}\bigg]\bigg\}\\\notag
&=& \frac{1}{N}\sum_{u=0}^{2q-1}\exp\bigg\{j2\pi\bigg[-u^2(\frac{p}{q}-1)\frac{N}{4q}-u\tau\bigg]\\\notag
&&+j\pi u\bigg[3(\frac{p}{q}-1)\bigg]\bigg\}
\end{eqnarray}
Without the loss the generality, we let $p > q$. Because $p$ and $q$ are both the powers of 2, $p/q$ is an even integer. Because $N > 2q$ and $N$ is the power of 2, $N/(4q)$ is an integer. So the term $[-u^2(p/q-1)N/(4q)-u\tau]$ is an integer as well. Thus,
\begin{eqnarray}
|C^{(p,q)}(\tau)|^2 = \frac{1}{N}\sum_{u=0}^{2q-1}\exp\bigg\{j\pi u \bigg[3(\frac{p}{q}-1)\bigg]\bigg\}
\end{eqnarray}
Because $p/q$ is even, the term $[3(p/q-1)]$ is always odd. Thus,
\begin{eqnarray}
|C^{(p,q)}[\tau]|^2
&=& \frac{1}{N}\sum_{u=0}^{2q-1}(-1)^u \\\notag
&=& 0
\end{eqnarray}
Thus,
\begin{eqnarray}
C^{(p,q)}[\tau] = 0, \, \tau=0,1,\dots,N-1
\end{eqnarray}

\section{The Cramer-Rao Bound}
The $\mathbf{r}_m$ in (\ref{model_matrix}) is Gaussian distributed as
\begin{equation}
\label{rmg}
\mathbf{r}_m \sim N\bigg(\mathbf{\mu(\mathbf{h}_{i,m})},2\sigma_m^2\mathbf{I}_N\bigg)
\end{equation}
where $\mathbf{\mu(\mathbf{h}_{i,m})}= \mathbf{S}_i\mathbf{h}_{i,m}+\sum_{j\neq i}\mathbf{S}_j\mathbf{h}_{j,m}$.
The Fisher information matrix can be obtained as
\begin{eqnarray}
\label{fisher}
J(\mathbf{h}_{i,m}) &=& \bigg(\frac{\partial \mathbf{\mu(\mathbf{h}_{i,m})}}{\partial \mathbf{h}_{i,m}}\bigg)^H(2\sigma_m^2\mathbf{I}_N)^{-1}\bigg(\frac{\partial \mathbf{\mu(\mathbf{h}_{i,m})}}{\partial
\mathbf{h}_{i,m}}\bigg)\\\notag
&=&\frac{1}{2\sigma_m^2}\mathbf{S}_i^H\mathbf{S}_i\\\notag
&=&\frac{1}{2\sigma_m^2}\mathbf{I}_L
\end{eqnarray}
So the Cramer-Rao Bound is
\begin{equation}
\label{crb}
\text{var}(\mathbf{\hat{h}}_{i,m})\geq2L\sigma_m^2
\end{equation}

%


%
%
%
%
%
%
%
%
%
%
\bibliographystyle{IEEEtran}
\bibliography{IEEEabrv,channel}
\end{document}